\documentclass[aps,showpacs,twocolumn,superscriptaddress]{revtex4-1}
\usepackage[pdftex]{graphicx}
\usepackage[english]{babel}
\usepackage[pdftex]{graphicx}
\usepackage{placeins}

\graphicspath{{./pics/}}

\begin{document}

\title{Pulse shaping using dispersion-engineered difference frequency generation}

\author{M. Allgaier}
\affiliation{Integrated Quantum Optics, Applied Physics, University of Paderborn, 33098 Paderborn, Germany}
\author{V. Ansari}
\affiliation{Integrated Quantum Optics, Applied Physics, University of Paderborn, 33098 Paderborn, Germany}
\author{J. M. Donohue}
\affiliation{Integrated Quantum Optics, Applied Physics, University of Paderborn, 33098 Paderborn, Germany}
\affiliation{Institute for Quantum Computing, University of Waterloo, 200 University Ave. West, Waterloo, Ontario, Canada, N2L 3G1}
\author{C. Eigner}
\affiliation{Integrated Quantum Optics, Applied Physics, University of Paderborn, 33098 Paderborn, Germany}
\author{V. Quiring}
\affiliation{Integrated Quantum Optics, Applied Physics, University of Paderborn, 33098 Paderborn, Germany}
\author{R. Ricken}
\affiliation{Integrated Quantum Optics, Applied Physics, University of Paderborn, 33098 Paderborn, Germany}
\author{B. Brecht}
\affiliation{Integrated Quantum Optics, Applied Physics, University of Paderborn, 33098 Paderborn, Germany}
\author{C. Silberhorn}
\affiliation{Integrated Quantum Optics, Applied Physics, University of Paderborn, 33098 Paderborn, Germany}

\date{\today}

\begin{abstract}
The temporal-mode (TM) basis is a prime candidate to perform high-dimensional quantum encoding. Quantum frequency conversion has been employed as a tool to perform tomographic analysis and manipulation of ultrafast states of quantum light necessary to implement a TM-based encoding protocol. While demultiplexing of such states of light has been demonstrated in the Quantum Pulse Gate (QPG), a  multiplexing device is needed to complete an experimental framework for TM encoding. In this work we demonstrate the reverse process of the QPG. A dispersion-engineered difference frequency generation in non-linear optical waveguides is employed to imprint the pulse shape of the pump pulse onto the output. This transformation is unitary and can be more efficient than classical pulse shaping methods. We experimentally study the process by shaping the first five orders of Hermite-Gauss modes of various bandwidths. Finally, we establish and model the limits of practical, reliable shaping operation.
\end{abstract}

\maketitle

High-dimensional encoding can potentially increase the security of quantum communication protocols as well as the information capacity of a single photon \cite{rohde_information_2013,hayat_multidimensional_2012}. Orbital angular momentum (OAM) has been proposed as such a basis for high-dimensional encoding \cite{leach_quantum_2010} but is inherently incompatible with existing telecommunication fiber networks. Temporal modes (TMs) of ultrafast pulses of light are a viable, fiber-compatible alternative to OAM, owing to their spatially single-mode nature \cite{brecht_photon_2015}. The core of the TM framework is the Quantum Pulse Gate (QPG), a non-linear optical device based on dispersion-engineered quantum frequency conversion in non-linear waveguides \cite{eckstein_quantum_2011}. The QPG has been shown to perform efficient sorting (i.e. demultimplexing) of the orthogonal but field-overlapping modes \cite{reddy_temporal_2013, brecht_demonstration_2014, ansari_temporal-mode_2017, ansari_tomography_2018, manurkar_multidimensional_2016, reddy_engineering_2017, reddy_temporal-mode-selective_2017}, as well as state manipulation and purification \cite{ansari_tomography_2018}, photon subtraction \cite{averchenko_nonlinear_2014,ra_tomography_2017} and noise suppression \cite{shahverdi_quantum_2017}. Quantum light in TM basis can be directly generated using an adapted parametric down-conversion source \cite{ansari_heralded_2018}.  For two-dimensional states, reshaping (i.e. modal rotation) has been explored \cite{manurkar_programmable_2017}. However, an independent TM multiplexing device capable or arbitrary TM shaping and reshaping of higher order modes such as the Quantum Pulse Shaper (QPS) described in Ref. \cite{brecht_quantum_2011} has not been demonstrated, neither on the single photon level, nor classically. Such a device, together with a QPG, could perform rotations between TMs. The process can in principle be very efficient, contrary to established classical pulse shaping methods in the spectral \cite{weiner_ultrafast_2011} and time domain \cite{rogers_nanosecond_2016}.

In this work we experimentally demonstrate a difference frequency generation (DFG) based pulse shaper and study its performance using coherent light. We verify successful shaping of the converted light into the first five orders of Hermite-Gauss modes using spectral intensity measurements. We assess the process' shaping accuracy by scanning the bandwidth of the desired spectrum and model experimental imperfections for comparison. We thus establish a range of working parameters for such a pulse reshaping device.

To design a DFG pulse shaping device such as the QPS, one has to first revisit the working principle behind the QPG. Key to the QPG device is its unique group-velocity relationship: By matching the group velocity of input and pump field in a sum-frequency generation (SFG) process, the conversion efficiency is directly proportional to the temporal overlap of the two fields, thus allowing to selectively converting field-orthogonal modes. This group-velocity matching has been achieved in two different ways: Using a almost degenerate type-0 non-linear process \cite{reddy_engineering_2017,manurkar_multidimensional_2016}, or by compensating for the waveguide's dispersion with material birefingence using a type-II process \cite{brecht_quantum_2011}. The latter has the prospect of better background suppression, resulting in signal-to-noise ratio sufficient for operating on quantum light \cite{ansari_tomography_2018}. This particular type-II implementation addresses inputs in the telecom band around 1550\,nm and outputs them in the visible range around 557\,nm . A QPS should ideally work in the opposite direction, thus enabling us to reconvert the output of a QPG to allow for TM rotations. Therefore, we propose to employ the reverse process of the QPG, i.e. difference-frequency generation (DFG). Here, using a type-II process is especially advantageous over type-0: The single photon output at 1550\,nm for a type-0 process can be separated from its pump field only by a few nanometers, and therefore be polluted by Raman-scattered photons. The process is  implemented in periodically poled Lithium Niobate waveguides. Employing the exact reverse process for SFG and DFG, i.e. same material, wavelengths and polarizations, allows the usage of the same waveguide structure and poling period. For group-velocity matched SFG, highly efficient bandwidth compression has already been demonstrated \cite{allgaier_highly_2017}. In an analog fashion, pulse shaping implemented in the DFG process can also be highly efficient, and even bandwidth expansion is possible, with potential applications in interfacing with narrowband photons.

\begin{figure}
\centering
\includegraphics[width=0.23\textwidth]{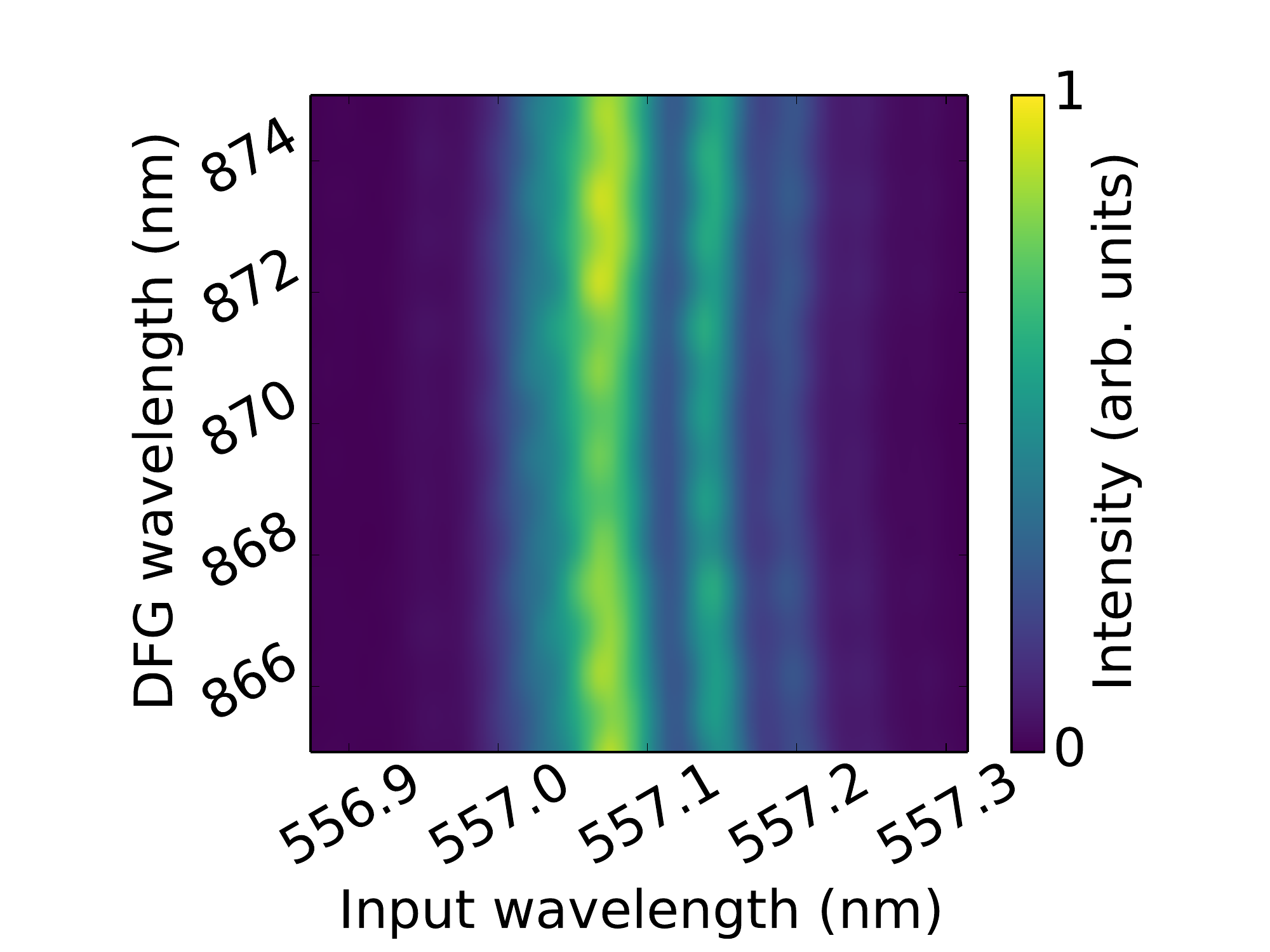}
\includegraphics[width=0.235\textwidth]{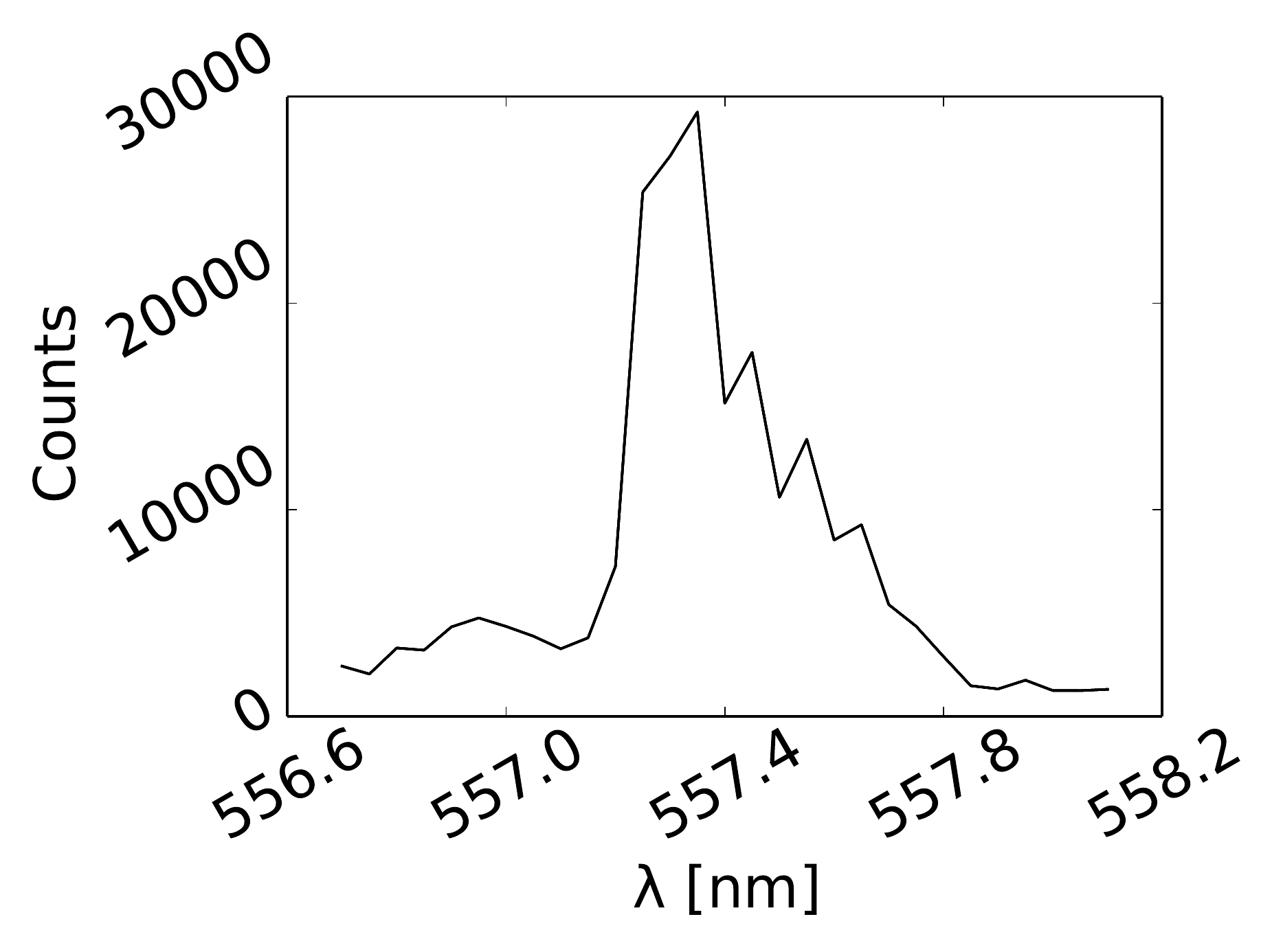}
\caption{Left: DFG phasematching derived from the measured SFG phasematching. Right: Phasematching scan at a pump wavelength of 1550\,nm}
\label{pm}
\end{figure}

\begin{figure}
\centering
\includegraphics[width=0.5\textwidth]{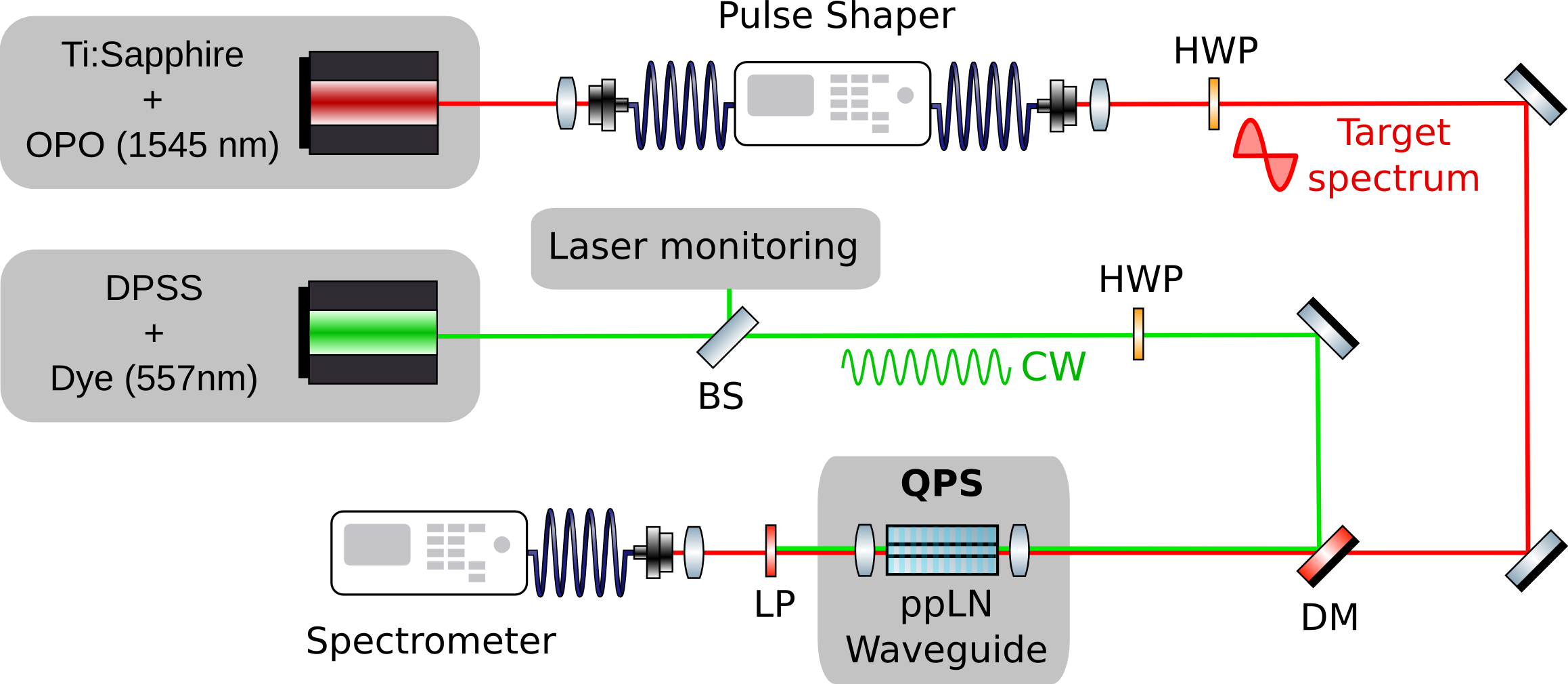}
\caption{The experimental setup for device characterization. OPO: Optical parametric oscillator, DPSS: Diode pumped solid state laser, HWP: Half wave plate, Ti:ppLN: periodically poled Lithium Niobate, LP: Long pass filter}
\label{setup}
\end{figure}

For the DFG device presented here, a central pump wavelength of 1550\,nm is chosen in order to convert narrowband light at 557\,nm to 850\,nm. This interchange of pump and output wavelength is possible due to the matched group-velocity, and the necessary lasers and classical pulse shapers were already available to the authors. From the SFG phasematching measured in a QPG configuration \cite{allgaier_highly_2017}, we derive the corresponding DFG phasematching with the accordingly different input and output wavelength and show it in Figure \ref{pm}. Directly scanning the phasematching in the DFG configuration was not possible due to poor repeatability of the 557\,nm laser's tuning, but a single scan through the phasematching verifies the derived phasematching.
The experimental setup is shown in Figure \ref{setup}. To generate the input light we rely on a diode-pumped solid state laser emitting at 514\,nm to pump a standing-wave continuous wave dye laser. Rhodamine-560 is employed as a laser dye to generate the necessary wavelength of 557\,nm. The laser's emission bandwidth is typically of the order of 5 GHz. A pulsed laser at 550\,nm was not available to the authors. The pump pulses are generated with a cascade of a Ti:Sapphire oscillator operating at 80\,MHz repetition rate, and an optical parametric oscillator emitting pulses with a central wavelength of 1550\,nm. The light is coupled to standard SMF-28 fibers and fed through a fiber-coupled spatial light modulator-based pulse shaper with a resolution of 10\,GHz over the entire telecom C-band. The pulses are combined with the 557\,nm input light on a dichroic mirror and coupled into a 27\,mm long, home-made Titanium-indiffused Lithium Niobate waveguide with a poling period of 4.4\,\(\mu\)m. Input and pump pulses are orthogonally polarized since we are employing a type-II process. The converted light is separated from the input light and coupled to another single mode fiber. We employ a single-photon sensitive spectrometer with a resolution of 0.05\,nm at 870\,nm, provided by a 1200\,lines/mm grating. For analysis, we calculate the expected DFG spectrum from the spectrum programmed on the pulse shaper. This is particularly simple due to the continuous wave input. We compare this spectrum to the measured one by means of an overlap integral. Unfortunately, with the available input and pump power, the generated pulses were too weak to characterize their spectral phase using classical pulse characterization techniques.

\begin{figure}
\centering
\includegraphics[width=0.5\textwidth]{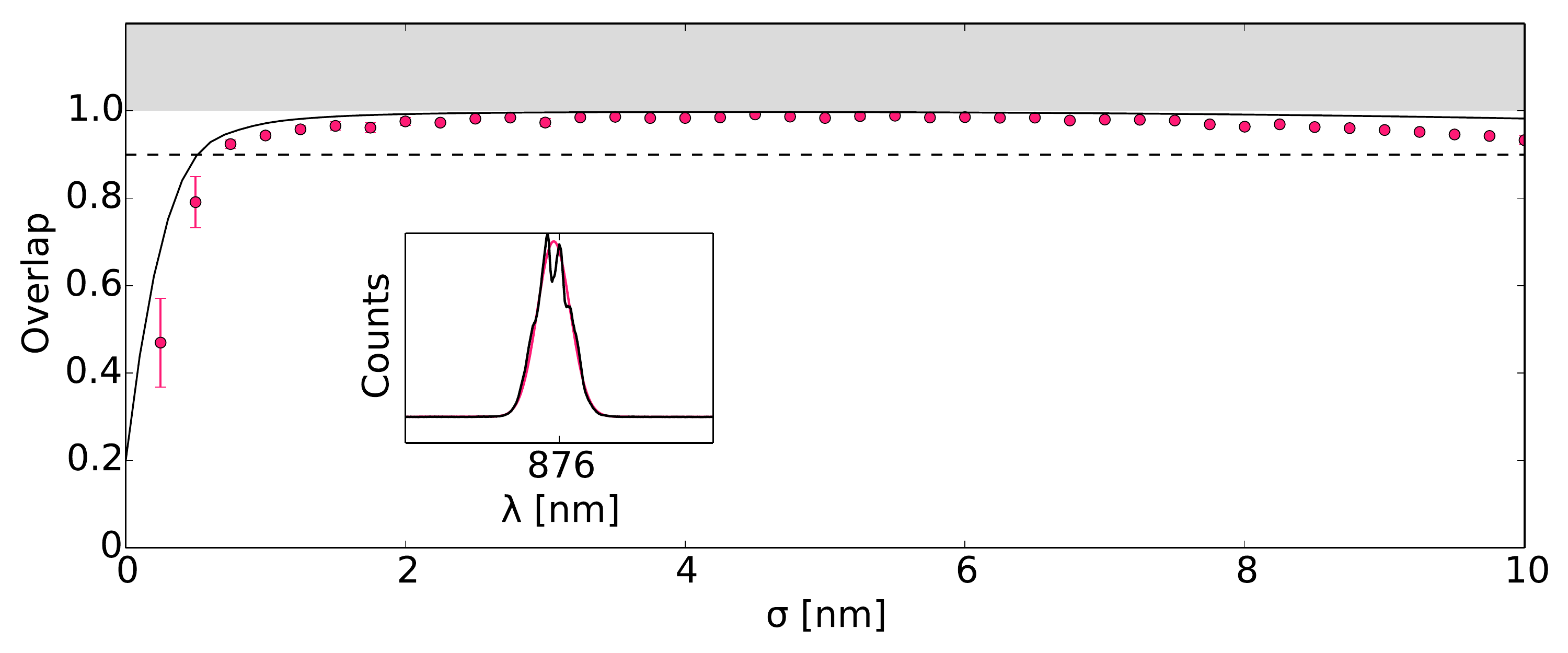}
\includegraphics[width=0.5\textwidth]{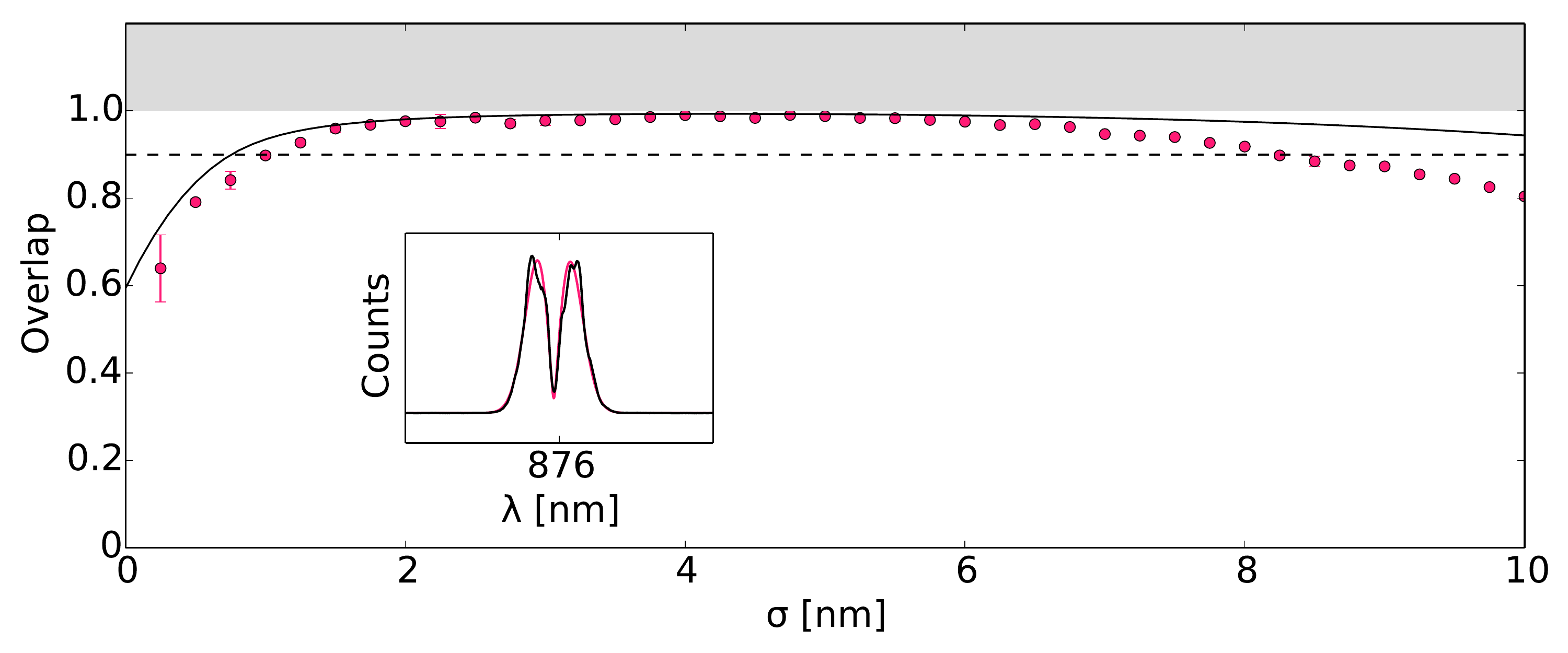}
\includegraphics[width=0.5\textwidth]{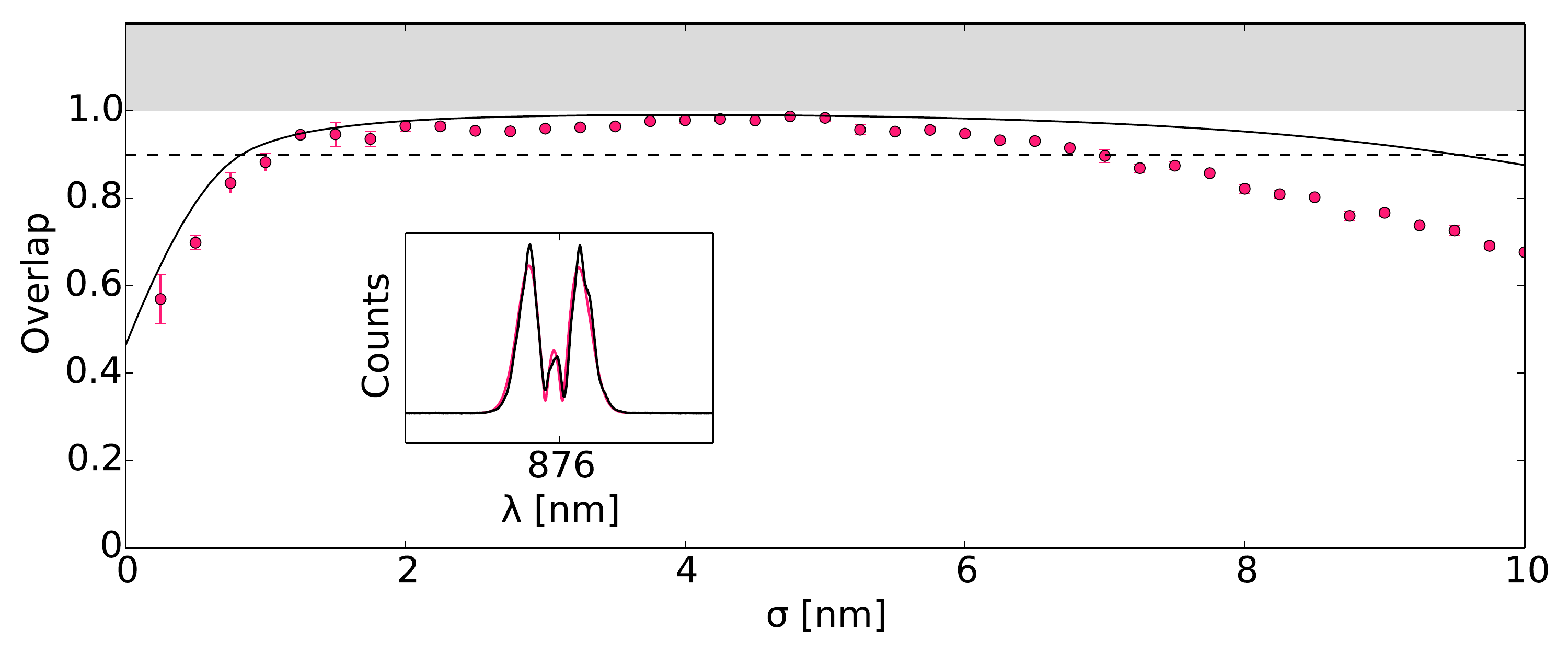}
\includegraphics[width=0.5\textwidth]{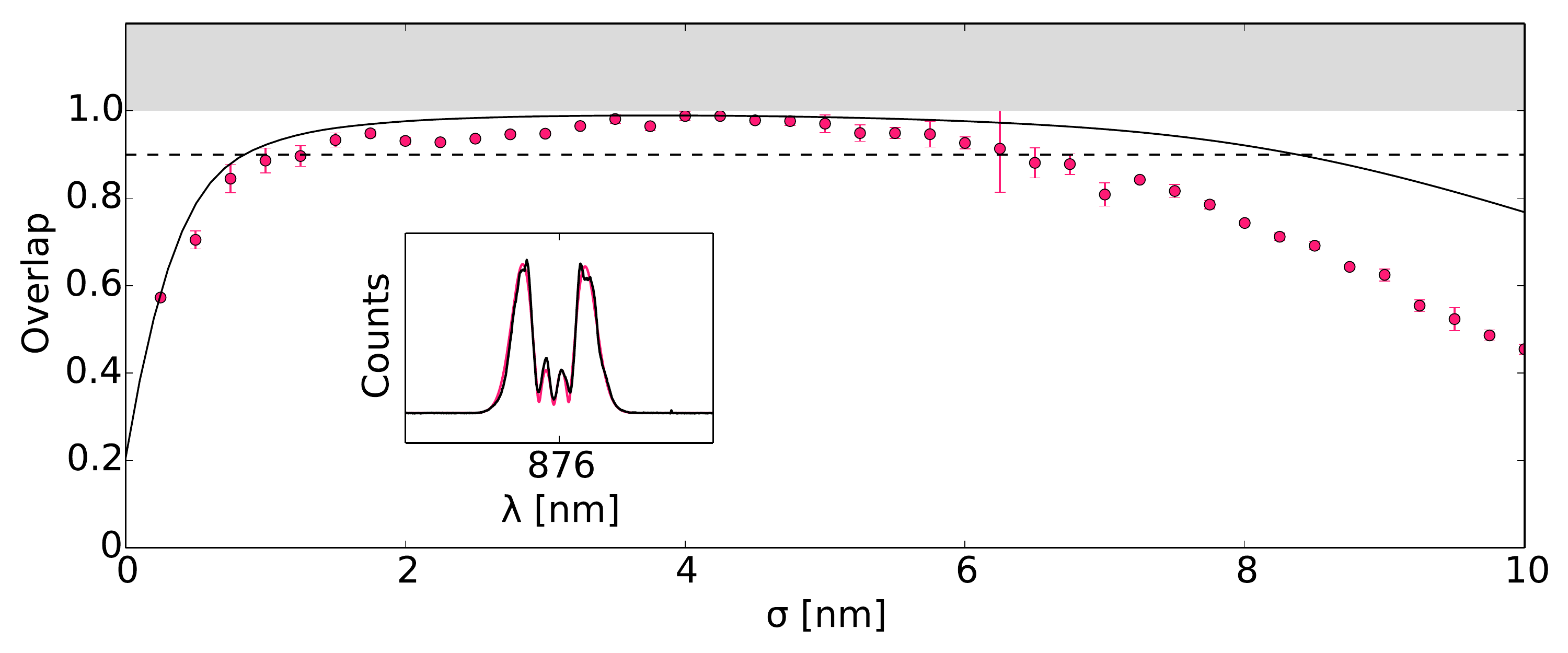}
\includegraphics[width=0.5\textwidth]{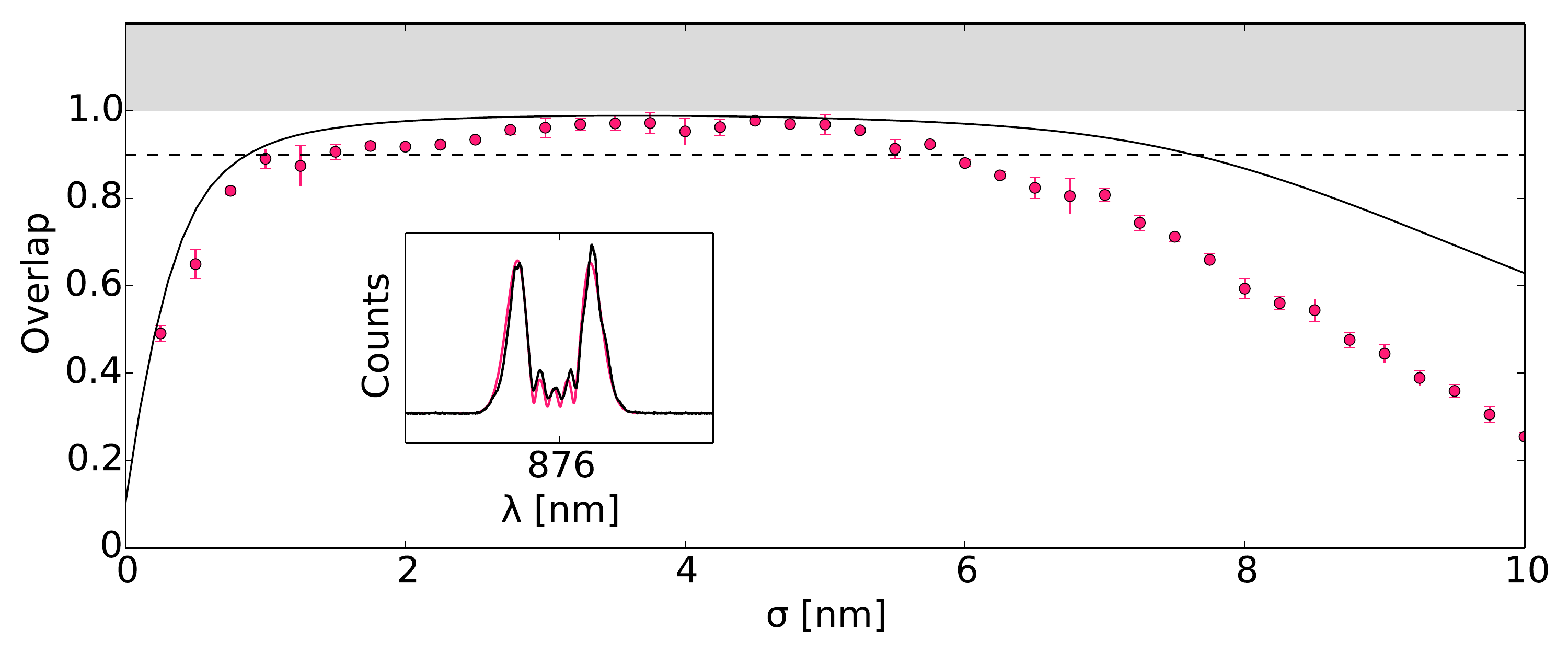}
\caption{Overlap between the programmed and measured Hermite-Gauss spectra for the first five modes. The dashed lines indicate 95\,\% overlap. Solid lines indicate the overlap between the programmed and modeled spectra, which include experimental limitations. The inserts show the programmed (magenta) and measured (black) spectra for each mode and a bandwidth of 5\,nm.}
\label{overlaps}
\end{figure}

Light produced by parametric down-conversion sources naturally decomposes into the Hermite-Gauss basis \cite{walborn_generalized_2012}. Since the device presented here is intended for use in the QPG framework, we choose exactly this basis. The spectral envelope of the pulses read:
\begin{equation}
HG_n(\lambda) \ = \ \frac{H_n(\lambda-\lambda_0)}{N_n} \ \cdot \ e^{\frac{(\lambda-\lambda_0)^2}{2\sigma^2}}
\end{equation}
where \(H_n\) denotes the Hermite polynomial of order \(n\), \(\lambda_0\) is the central wavelength, and \(\sigma\) denotes the base Gaussian's bandwidth in the following, although the actual spectral spread will be higher for higher-order modes. \(N\) normalizes the Hermite-Gauss function. We scan the bandwidth of the underlying Gaussian from 0.25\,nm to 10\,nm in steps of 0.25\,nm, and perform 8 measurements with 4 seconds integration time each for every set of parameters. The standard deviation over the mean value of the 8 measurements is used to generate error bars. Thus, for every Hermite-Gauss order 320 measurements are taken, this task is repeated for the first 5 orders of Hermite-Gauss modes over the course of half a day. For every bandwidth \(\sigma\), we calculate the overlap between target and measured spectrum:
\begin{equation}
OL = \frac{\left(\int S(\lambda)T(\lambda)d\lambda\right)^2}{\int S^2(\lambda)d\lambda\cdot \int T^2(\lambda)d\lambda}
\end{equation}
where \(S(\lambda)\) and \(T(\lambda)\) are the measured and programmed target spectra, respectively. The results are displayed in Figure \ref{overlaps}.

The overlap between the measured programmed spectra are displayed in magenta. The achieved overlap is above 90\,\% over a large bandwidth range from about 1 to 6\,nm for all five modes. The best overlap at over 95\,\% is achieved for bandwidths between roughly 3.5 and 5\,nm. It is noteworthy that a higher order Hermite-Gauss mode of the same nominal bandwidth occupies a wider spectrum, since the Gaussian is scaled by the Hermite polynomial. This causes the steeper decline of the overlap for wider bandwidths and higher orders. To better understand these results, we modeled the experimental limitations of the current setup. The overlap between these modeled spectra and the programmed ones are again calculated and displayed as solid lines. The model contains two classes of contributions.

\begin{figure}
\centering
\includegraphics[width=0.28\textwidth]{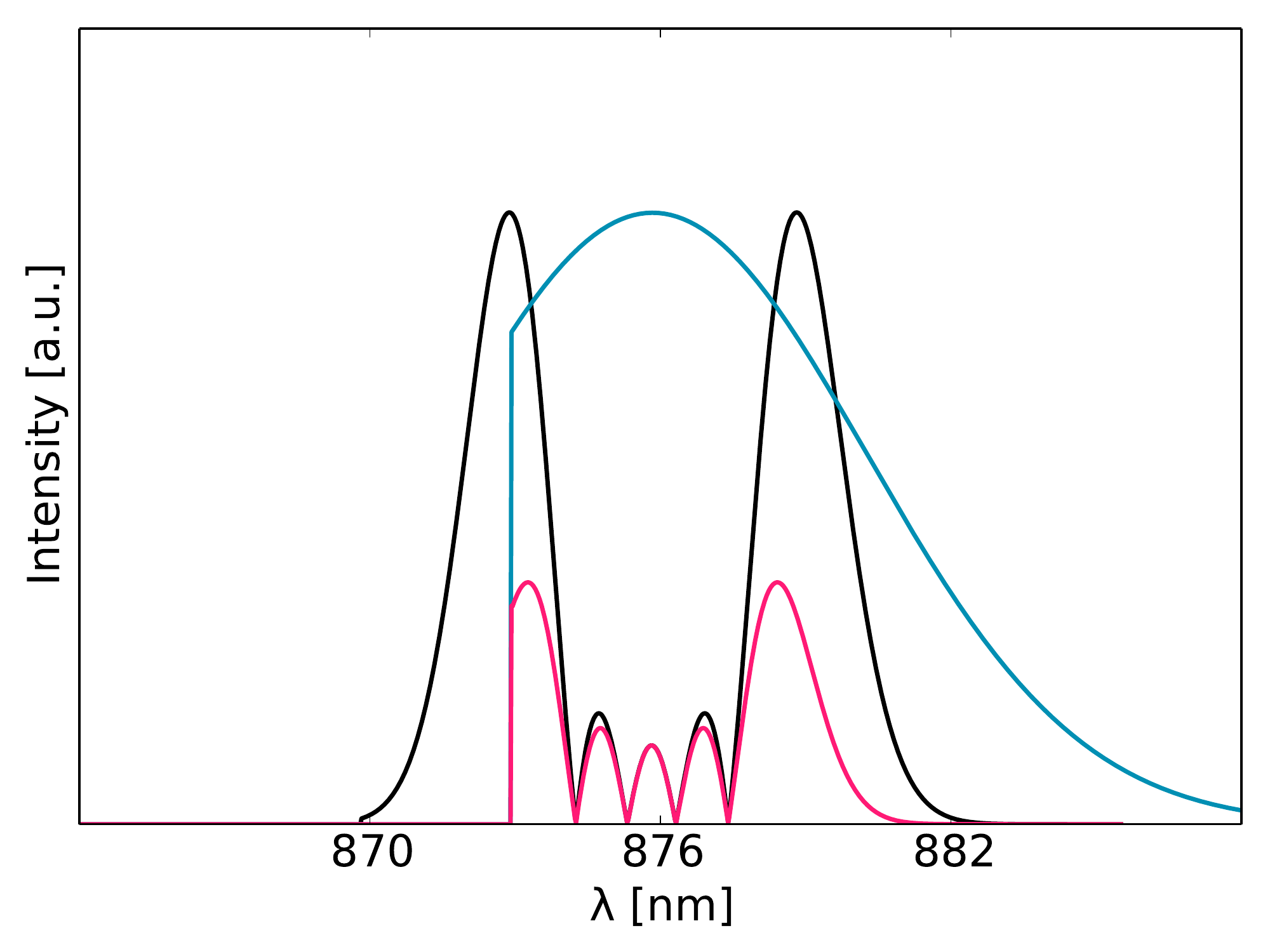}
\caption{Sketched programmed spectrum (black), available pump spectrum with limited shaper range (blue) and resulting actual shaped spectrum (magenta) for a forth order Hermite-Gauss mode.}
\label{cutmodel}
\end{figure}

First, we account for insufficient available bandwidth at 1550\,nm, which effectively narrows down the shaped spectrum. This includes three contributions: The available bandwidth from the pump laser, the available phasematching bandwidth, and the range of the pulse shaper. While the first two contributions result in a combined Gaussian spectrum of 10\,nm FWHM, with which the programmed spectrum is multiplied. Figure \ref{cutmodel} shows this effect for the 4th order mode with a bandwidth of 10\,nm. The black line is the programmed spectrum. However, it can only be carved within the operating bandwidth of the pulse shaper, and only from the Gaussian spectrum with limited bandwidth (shown in blue), which is why the actual shaped spectrum (magenta) looks different from the programmed one. The difference is mostly visible in the outer parts of the spectrum, which is why only large bandwidths and higher-order modes are affected.

%This scales down the wings of the spectrum. The contribution from the pulse shaper results in the spectrum simply being cut at a certain point (below 1530 and above 1565\,nm). The latter is only visible at higher orders and large bandwidths at the edge of the distribution shown here. These effects cause the decline of the theory curves at large bandwidths and are more apparent for higher orders due to their larger effective bandwidth (which is larger since the Gaussian, nominal bandwidth is scaled up by the polynomial contribution).

Second, we model  convolution effects. These are due to limited resolution of the employed pulse shaper (10\,GHz) and spectrometer (20\,GHz), as well as the non-negligible bandwidth of the dye laser (5\,GHz). The programmed spectrum is convoluted with a Gaussian to account for these effects. Another such effect stems from multimodeness of the dye laser, which experiences a certain degree of mode competition. The two modes of 5\,GHz bandwidth are estimated to be about 0.1\,nm apart. We account for this by adding up twp spectra of the same separation. We estimate the ratio to be 1:1, since the mode competition takes place on the order of seconds, whereas the measurement for each bandwidth takes 32 seconds, thus averaging sufficiently over both mode contributions. These effects  blur the spectrum, and lead to diminished overlap for small bandwidths.

%Another imperfection in the measured spectra is noise. We modeled this as well by adding normally distributed noise to the model spectra, and found it to diminish the overlap for all bandwidths, however, an unreasonably large amount of noise was necessary to produce a significant effect. Noise is thus excluded from the model.

It is apparent that the model does not fit well the large bandwidths, while the qualitative trend is still reproduced. We attribute this to instabilities in the spectra produced by the optical parametric oscillator. It is possible to observe fringing effects in the pump spectrum in front of the pulse shaper, manifesting as small features in the spectra. In addition, the pump spectrum in front of the pulse shaper is not exactly Gaussian and exhibits some degree of asymmetry. These features change over time on a scale of 10 seconds and cause more pronounced deviations for more complex and wider spectra. 

\begin{figure}
\centering
\includegraphics[width=0.5\textwidth]{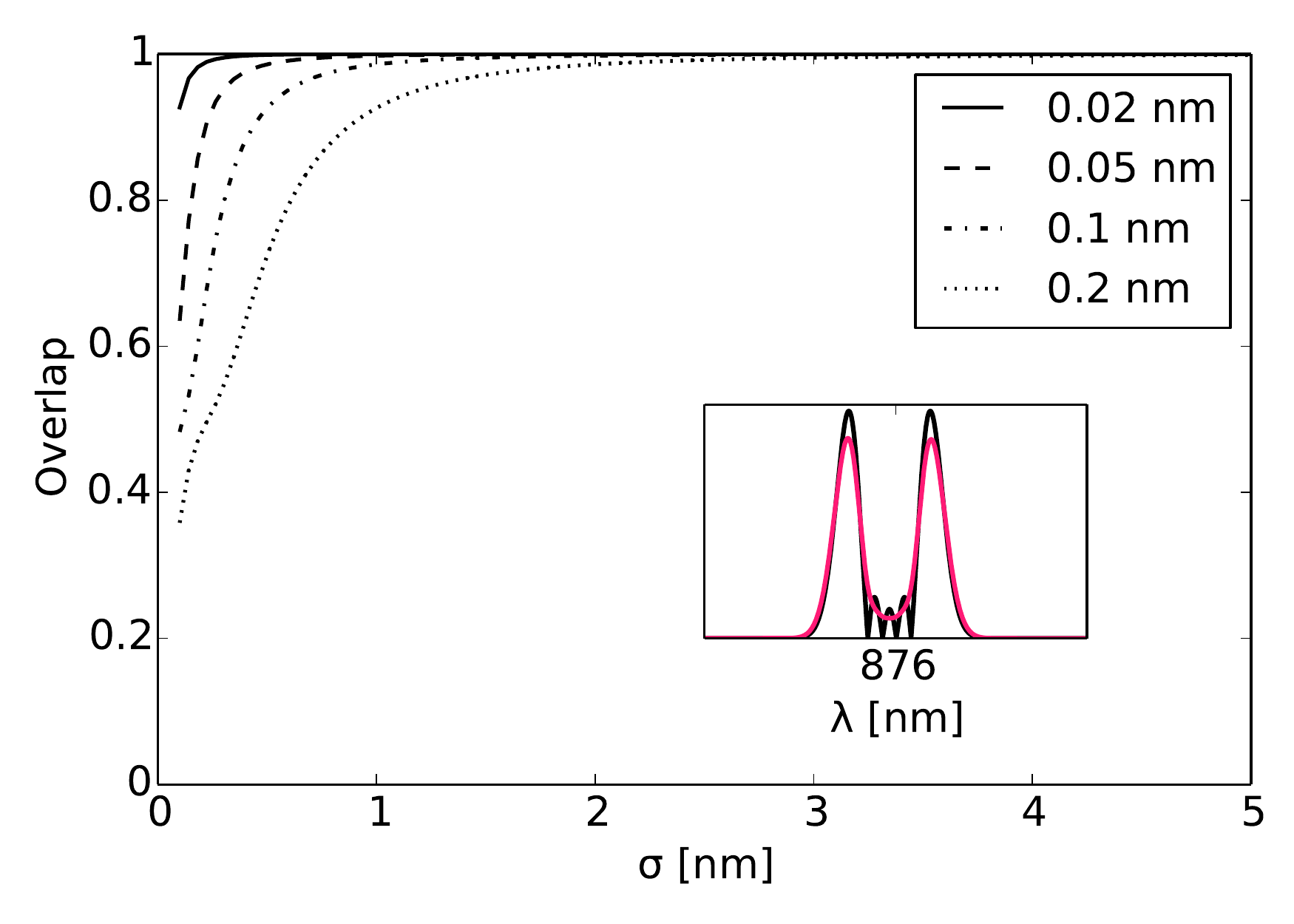}
\caption{Overlap between the programmed and modeled spectra of the fourth Hermite-Gauss order for various phasematching bandwidths. The inset shoes the target (black) and model (magenta) spectrum for a phasematching bandwidth of 0.2\,nm and a mode bandwidth of 1.8\,nm.}
\label{outlook}
\end{figure}

The individual contributions to the model are treated in more detail in the supplementary materials. From the achieved overlaps we conclude that the device works in principle, with some constraints imposed by the current experimental setup. However, it is important to dissect which of the imperfections are fundamental to the device, and which are only caused by auxiliary equipment such as the lasers. First, the spectrometer resolution is not a fundamental restriction for the device, since it only influences how well the shaped spectra can be characterized, not how well they are shaped. The effects imposed by the pulse shaper are device dependent. This leaves the 557\,nm input bandwidth and phasematching bandwidth as ultimate limits to device performance. Therefore, we model the influences of these ultimate limitations on the proposed pulsed input device. The input bandwidth should always be chosen smaller or equal the phasematching bandwidth, or otherwise the spectrum will be cut and effectively filtered once more
We now assume a flat spectral intensity and phase of the pump laser spectrum, as well as a flat response of the pulse shaper. The pulse shaper's resolution of 10\,GHz will be small compared to the phasematching bandwidth and neglected. Still, the non-zero input bandwidth results in a convolution effect just like the ones discussed above.
%The phasematching bandwidth along the output axis is mostly limited by the phasematching's curvature and would thus not strongly depend on the bandwidth.
Using those benchmark numbers, calculations identical to the model already presented were prepared to simulate the effect of the input and phasematching bandwidth on the quantum device performance. Since the highest order mode used is subject to the strongest limitations, we only show the results for the 4th order Hermite-Gauss mode in Figure \ref{outlook}. It can be seen that for a more narrow phasematching bandwidth than the one for the waveguide used in this work, shaping bandwidths under 1\,nm is certainly possible. This would require longer waveguides. This is highly desirable in the light of optical fiber dispersion and spectral information density. At the same time, high shaping fidelity for small bandwidth features would also allow us to shape higher order modes efficiently. The inset shows the target and model spectrum for a mode bandwidth of 1.8\,nm and a phasematching bandwidth of 0.2\,nm. The blurring effect on the central features can clearly be observed. This is the source of the diminished overlap.

In conclusion, we have shown the classical characterization of a DFG pulse shaper and show successful reshaping of the input light into Hermite-Gauss pulses of a broad substantial range of bandwidths. From the theoretical model of the experimental imperfections we draw the conclusion that a highly functional device for pulsed operation can be implemented using the current waveguide technology.

\FloatBarrier

\section*{Acknowledgement}
This work was funded by the DFG through the SFB TRR 142.

%group velo matching, contrast to QPG, reverse process = DFG, useful wavelength combinations, different choice due to technical reasons in the lab.

\bibliographystyle{unsrt}
%\bibliography{own,telecom,memories,bandwidth,theory,materials,streak,quantumcomm}
\bibliography{quantumcomm,Telecom,bandwidth,spectral,theory}

\end{document}